\documentclass[aps,preprint,showpacs,showkeys,preprintnumbers,eqsecnum,amsmath,amssymb]{revtex4}

\usepackage{dcolumn}
\usepackage[dvips]{epsfig}

\begin{document}
\title{\bf Particle Acceleration on the Background of the Kerr-Taub-NUT Spacetime }

\author{
Changqing Liu$^{1,2}$, %\footnote{Electronic address: lcqliu2562@163.com}$,
Songbai Chen$^{1}$, %\footnote{Electronic address: csb3752@163.com}$,
Chikun Ding$^{2}$, %\footnote{Electronic address: dingchikun@163.com}
and Jiliang  Jing\footnote{Corresponding author, Email:
jljing@hunnu.edu.cn}$^{1}$ %\footnote{Electronic address:jljing@hunnu.edu.cn}
}

\affiliation{1) Department of Physics, and Key Laboratory of Low Dimensional Quantum Structures \\ and Quantum Control of Ministry of Education, Hunan Normal
University, \\ Changsha, Hunan 410081, People's Republic of China}

\affiliation{2) Department of Physics and Information Engineering, \\
Hunan Institute of Humanities Science and Technology,\\ Loudi, Hunan
417000, P. R. China}

\begin{abstract}

We study the collision of two particles  with the different rest
masses moving in the equatorial plane of a Kerr-Taub-NUT spacetime
and get the center-of-mass (CM) energy for the particles. We find
that the CM energy depends not only on the rotation parameter, $a$, but also on the NUT charge  of the Kerr-Taub-NUT  spacetime, $n$.
Especially, for the extremal Kerr-Taub-NUT spacetime, an unlimited CM energy can
be approached if the parameter $a$ is in the range $[1,\sqrt{2}]$,
which is different from that of the Kerr and Kerr-Newman black
holes.
\end{abstract}


\maketitle

\newpage
\section{Introduction}
Recently, Banados, Silk and West (BSW)  \cite{BSW2009} studied the
collision of two particles near a rotating black hole and found that
CM energy $E_{\rm cm}$ of the two particles moving along the
equatorial plane can be arbitrarily high in the limiting case of
maximal black hole spin. This means that the extremal rotating black
hole could be regarded as a Planck-energy-scale collider, which
might bring us some visible signals from ultra high energy
collisions, such as, dark matter particles. Thus, the BSW mechanism
about the collision of two particles near a rotating black hole has
been attracted much attention in the recent years. In Refs.
\cite{Berti_etal2009,Jacobson_Sotiriou2010}  Berti et al pointed out that the arbitrarily high CM energies  $E_{\rm
cm}$ for a Kerr black hole might not be achievable in nature due to
the astrophysical limitation,
 such as the maximal spin \cite{Thorne}, gravitational radiation.
Subsequently, Lake \cite{ Lake} investigated the CM energy of
the collision occurring at the inner horizon of the non-extremal Kerr
black hole and found the CM energy is limited \cite{ Lake}. Grib and Pavlov
\cite{Grib1,  Grib2, Grib_Pavlov2010_Kerr, Grib_Pavlov20102052}
argued that the CM energy $E_{\rm cm}$ for two particles collision
can be unlimited even in the non-maximal rotation if one considers
the multiple scattering, and they also evaluated extraction of
energy after the collision. The collision in the innermost stable
circular orbit was considered in Ref. \cite{Harada}. The similar BSW
mechanism had also been found in other kinds of black holes, e.g.
Stringy Black Hole \cite{Wei2}, Kerr-Newman black
holes \cite{WLGF2010} and Kaluza-Klein Black Hole \cite{Mao}. In
Refs. \cite{Zaslavskii2010_rotating,
Zaslavskii2010_charged,Zaslavskii3}, the author elucidated the
universal property of acceleration of particles for rotating black
holes and try to give a general explanation of this BSW mechanism
for the general rotating black holes. The BSW mechanism stimulated
some implications concerning the effects of gravity generated by
colliding particles in Ref. \cite{Kimura} and the emergent flux from
particle collision near the Kerr black holes \cite{bsw1010}.

Another interesting stationary axisymmetric object is the Kerr-Taub-NUT
(KTN) spacetime \cite{DemNew,Mil}, which is an important solution of
Einstein-Maxwell equations for electro-vacuum spacetime possessing
with gravitomagnetic monopole and dipole moments. Besides the mass
$M$ and the rotation parameter $a$, the KTN spacetime carries with the
NUT charge, $n$, which plays the role of a magnetic mass inducing a
topology in the Euclidean section. The presence of the NUT charge
brings this spacetime some special spacetime structure, such as,
Misner singularity. The KNT spacetime serves as an attractive example of spacetimes
with asymptotic non-flat structure for exploring
various physical phenomena in general relativity e.g. gravitomagnetism \cite{bini}. The
KNT solutions  representing relativistic
thin disks \cite{Gonzalo} are of great astrophysical importance since they can be used as models for certain
galaxies, accretion disks, and the superposition of a black holes and a galaxy or an accretion disk as
in the case of quasars. The main purpose of this paper is to study the
collision of two particles  with the different rest masses in the
background of the KTN spacetime and to see what effects of the NUT
charge on the CM energy for the particles in the near-horizon
collision.

This paper is organized as follows. In Sec. II we derive briefly
particle orbits in the KTN  spacetime. In Sec. III, we
study the collision of two particles with the different rest masses
moving in the equatorial plane of the Kerr-Taub-NUT spacetime and
discuss the center-of-mass (CM) energy for the particles. Sec.IV is
devoted to a brief summary. We use the units $c=G=1$ throughout the
paper.

\section{Particle orbits in Kerr-Taub-NUT spacetime}

The metric of the KTN spacetime in the Boyer-Lindquist coordinates
can be expressed as \cite{DemNew,Mil}
\begin{eqnarray}
\label{metrica} d s^2
   &=-&\frac{1}{\Sigma}(\Delta -a^2\sin^2\theta)d t^2
    +\frac{2}{\Sigma}[\Delta \Xi -a(\Sigma +a \Xi)\sin^2\theta]dt
d\phi \nonumber \\
    &&+\frac{1}{\Sigma}[(\Sigma +a \Xi)^2\sin^2\theta-\Xi^2\Delta]d \phi^2
    +\frac{\Sigma}{\Delta}d r^2 +\Sigma d\theta^2,
\end{eqnarray}
with
\begin{eqnarray}
\Sigma = r^2 +(n+a\cos\theta)^2, \;\;\Delta = r^2-2Mr-n^2
+a^2,\;\;\Xi = a \sin^2\theta -2n\cos\theta,
\end{eqnarray}
where $M$, $a$ and $n$ are the mass, the angular parameter and the
NUT charge. The radius of the event horizon and
the Cauchy horizon of the KTN  spacetime are $r_{H,C} =M \pm
\sqrt{M^2-a^2+n^2}$, respectively, which are roots of the equation
$\Delta =0$. The existence of the horizons requires $a^2\leq
M^2+n^2$. The angular velocity of the KTN spacetime at the outer
horizon is
\begin{equation}
\Omega_{H}=\frac{a}{2(r_HM+n^2)}=\frac{a}{2(M^2+n^2+M\sqrt{M^2+n^2-a^2})}.
\end{equation}
In Boyer-Lindquist coordinates system, the timelike and axial Killing vectors
 are given by
$\xi^{a}=\left(\frac{\partial}{\partial t}\right)^{a}$ and
$\psi^{a}=\left(\frac{\partial}{\partial \phi}\right)^{a}$, respectively.
 With the help of the Killing vectors $\xi^{a}$
and $\psi^{a}$, we have
the following conserved quantities along a geodesic on the equatorial plane
\begin{eqnarray}
E&=&-g_{ab}\xi^{a}u^{b}=-u_{t}=\frac{r^2-2Mr-n^2}{n^2+r^2}u^{t}-\frac{a(\Delta-(a^2+n^2+r^2))}{r^2+n^2}u^{\phi},
\label{eq:e}\\
L&=&g_{ab}\psi^{a}u^{b}=u_{\phi}=\frac{a(\Delta-(a^2+n^2+r^2))}{r^2+n^2}u^{t}+\frac{(a^2+n^2+r^2)^2-a^2\Delta}{r^2+n^2}u^{\phi},
\label{eq:L}
\end{eqnarray}
where $u^b$ is the four velocity defined by $u^b=\frac{dx^b}{d\tau}$, $\tau$ is the proper time for timelike  geodesics. Moreover, we can introduce a new
conserved parameter $\kappa$ defined as
\begin{eqnarray}\label{eq:phidot111}
\kappa=g_{ab}u^{a}u^{b},
\end{eqnarray}
whose values are given
by $\kappa=-1,0,1$ corresponding to timelike geodesics, null geodesics, and spacelike geodesics, respectively.

With the help of Eqs. (\ref{eq:e}), (\ref{eq:L}) and (\ref{eq:phidot111}),  we can obtain
\begin{eqnarray}
u^{t}&=&\frac{1}{(n^2+r^2)\Delta}\Big(E\Big\{(n^2+r^2)^2+a^2[3n^2+r(2M+r)]\Big
\}-2aL(n^2+Mr)\Big ),
\label{eq:tdot}\\
u^{\phi}&=&\frac{1}{(n^2+r^2)\Delta}\Big
\{2aE(n^2+Mr)+L[-n^2+r(-2M+r)]\Big \},
\label{eq:phidot}\\
u^r&=&\pm\Big[-\frac{\Delta}{n^2+r^2}+\frac{1}{(n^2+r^2)^2}
\Big\{E^2[(a^2+n^2+r^2)^2-a^2\Delta]\nonumber\\
&+&2E^2L[-a(a^2+n^2+r^2)+a\Delta]-L^2(n^2-2Mr+r^2)\Big\}
\Big]^{\frac{1}{2}},\label{urr}
\end{eqnarray}
where the quantities $E$ and $L$ are the specific energy and angular
momentum of the particle, respectively. And then the radial equation
for the timelike particle moving along geodesics in the equatorial
plane is described by
\begin{equation}
\frac{1}{2}u^ru^r+V_{\rm eff}(r)=0, \label{eq:eom}
\end{equation}
with the effective potential
\begin{eqnarray}
V_{\rm eff}(r)
&=&\frac{\Delta}{2(n^2+r^2)}-\frac{1}{2(n^2+r^2)^2}\Big \{E^2[(a^2+n^2+r^2)^2-a^2\Delta]\nonumber\\
&+&2E^2L[-a(a^2+n^2+r^2)+a\Delta]-L^2(n^2-2Mr+r^2)\Big
\}\label{eq:effective_potential}.
\end{eqnarray}
The circular orbit of the particle is defined by
\begin{eqnarray}\label{veercondition}
V_{\rm eff}(r)=0,~~~~ \frac{dV_{\rm eff}(r)}{dr}=0.
\end{eqnarray}
Since $u^{t}>0$, the condition
\begin{eqnarray}
E\Big\{(n^2+r_H^2)^2+a^2[3n^2+r(2M+r)]\Big\}\ge
2aL(n^2+Mr),\label{cb1}
\end{eqnarray}
must be satisfied.
As $r\rightarrow r_{H}$ for the timelike particle, this condition reduce to
\begin{equation*}
E\ge\frac{aL}{2(n^2+r_HM)}= \Omega_{H}L
\end{equation*}

\section{CM energy of two particles in the Kerr-Taub-NUT spacetime}

In this section, we will study the CM energy for the collision of
two particles moving in the equatorial plane of the KTN spacetime.
Let us now consider two colliding particles with rest masses $m_1$
and $m_2$.  We assume that two particles 1 and 2 are at the same
spacetime point with the four momenta
\begin{equation*}
p_{i}^{a}=m_{i}u^{a}_{i},
\end{equation*}
where $p_{i}^{a}$ and $u_{i}^{a}$ are the four momentum and the four
velocity of particle $i$ ($i=1,2$). The sum of the above two momenta
is given by
\begin{equation*}
p_{\rm t}^{a}=p_{(1)}^{a}+p_{(2)}^{a}.
\end{equation*}
Then the CM energy $E_{\rm cm}$ of the two particles is given by
\begin{equation}
E_{\rm cm}^{2}=-p^{a}_{\rm t}p_{{\rm
t}a}=-(m_1u^{a}_{(1)}+m_2u^{a}_{(2)})(m_1u_{(1)a}+m_2u_{(2)a}).\label{eq:center-of-mass_energy}
\end{equation}
Due to $u^au_a=-1$, we obtain
\begin{eqnarray}
\label{ermcm} \frac{E_{\rm
cm}^{2}}{2m_1m_2}=\frac{m^2_1+m^2_2}{2m_1m_2}
-g_{ab}u^a_{(1)}u^b_{(2)},
\end{eqnarray}
which can be rewritten as
\begin{eqnarray}
\label{ermcm1} \frac{E_{\rm
cm}}{\sqrt{2m_1m_2}}=\sqrt{\frac{(m_1-m_2)^2}
{2m_1m_2}+(1-g_{ab}u^a_{(1)}u^b_{(2)})}.
\end{eqnarray}
In the case of $m_1=m_2$, the above the equation reduce the result
in Refs. \cite{BSW2009},\cite{Grib_Pavlov2010_Kerr}. On the background
metric (\ref{metrica}), substitute Eqs.~(\ref{eq:tdot}),
(\ref{eq:phidot}) and (\ref{urr}) into Eq.~(\ref{ermcm1}), the CM
energy of two particles in the KTN spacetime is shown by
\begin{eqnarray}
\frac{E^{2}_{\rm
cm}}{\sqrt{2m_1m_2}}=\sqrt{\frac{(m_1-m_2)^2}{2m_1m_2}+\frac{
A(r)-B(r)}{C(r)}}, \label{eq:ECOM_Ktn}
\end{eqnarray}
with
\begin{eqnarray}
A(r)&=&-2a(E_2L_1+E_1L_2)(n^2+r)+a^2\Big((1+3E_1E_2)n^2+r(r+E_1E_2
(2+r))\Big)\\\nonumber&-&(n^2+r^2)(n^2-E_1E_2n^2+2r-r^2-E_1E_2r^2)+L_1L_2(n^2-(r-2)r),
\label{eq:Fy}\\
B(r)&=&\sqrt{b_{1}(r)b_{2}(r)},
\label{eq:Gy}\\
b_{i}(r)&=&-4aE_iL_i(n^2+r)+L_i^2(n^2-(r-2)r)\\
\nonumber&+&(n^2+r^2)\Big((1+E^2_i)n^2+r(2+(E^2_i-1)r)\Big)
+a^2\Big((3E^2_i-1)n^2+r(E^2_i(2+r)-r)\Big),
\label{eq:giy} \\
C(r)&=&(n^2+r^2)(a^2-n^2+(2-r)r), \label{eq:Dy}
\end{eqnarray}
where $E_{i}$ and $L_{i}$ are the specific energy $E$ and the
angular momentum $L$ for particle $i$. Obviously, the result
confirms that the NUT charge $n$ indeed has
influence on the CM energy.

\subsection{Near-horizon collision in non-extremal KTN spacetime}
\label{subsec:subextreme}

We are now in the position to study the properties of the CM energy
(\ref{eq:ECOM_Ktn}) as the radius $r$ approaches to the horizon
$r_H$ of the non-extremal KTN spacetime. Note that both denominator
and the numerator
  of the
fraction $\frac{A(r)-B(r)}{C(r)}$ on the right-hand side of
Eq.~(\ref{eq:ECOM_Ktn}) vanishes at $r_H$. Using l'Hospital's rule
and taking into account $r_H^2-2r_H-n^2+a^2=0$, the value of
$E^{2}_{\rm cm}$ at $r_H$ is given by
\begin{equation*}
\frac{E^{2}_{\rm cm}(r\to r_H)}{\sqrt{2m_1m_2}}
=\sqrt{\frac{(m_1-m_2)^2}{2m_1m_2}+\frac{A(r)'-B(r)'}{C(r)'}}~\Bigg|_{r=r_H},
\end{equation*}
with
\begin{eqnarray*}
A'(r)|_{r=r_H}&=&-2a(E_2L_1+E_2L_2)+2(E_1E_2-1)n^2-2L_1L_2(r_H-1)\\\nonumber&+&4E_1E_2r_H+
2(1+3E_1E_2)n^2r_H+2(E_1E_2-1)r_H^2+2(1+E_1E_2)r_H^3, \\\nonumber
b_{i}'(r)|_{r=r_H}&=&
-4aE_iL_i+2(1+E_i^2)n^2-2L^2(r_H-1)\\\nonumber&+&4E_ir_H+2(3E_i^2-1)n^2r_H+2(1+E_i^2)r_H^2+
2(E_i^2-1)r_H^3, \\
C'(r)|_{r=r_H}&=& 2(r_H-1)(n^2+r_H^2), \\
B'(r)|_{r=r_H}&=&B\frac{1}{2}\left(
\frac{b_{1}'}{b_{1}}+\frac{b_{2}'}{b_{2}}\right).
\end{eqnarray*}
By implementing the calculation and taking the limit, we reach
\begin{eqnarray}
&&\frac{E_{\rm cm}(r\to r_H)}{2\sqrt{m_1m_2}}=
\Bigg\{\frac{(m_1-m_2)^2}{4m_1m_2}+1
\nonumber \\
&&+\frac{[(L_{H1}-L_{1})-(L_{H2}-L_{2})]^{2}
+(L_{H1}L_{2}-L_{H2}L_{1})^{2}\frac{2(n^2+1)-a^2-2\sqrt{1+n^2-a^2}}
{4(n^2+1+\sqrt{1+n^2-a^2})^2}}{4(L_{H1}-l_{1})(L_{H2}-L_{2})}
\Bigg\}^{\frac{1}{2}}. \label{eq:general_subextremal}
\end{eqnarray}
This is the formula for the CM energy of two particles along the
general geodesic orbits on the outer horizon. The critical angular
momenta $L_{Hi}~(i=1,2)$ can be written as
$L_{Hi}=\frac{E_i}{\Omega_H}=\frac{2E_{i}(n^2+r_H)}{a}$. In fact, as
we will prove in section ~\ref{subsec:extreme},
Eq.~(\ref{eq:general_subextremal}) is valid even in the extremal KTN
spacetime simply by taking the near-extremal limit
$a\to\sqrt{1+n^2}$. The necessary condition for obtaining an
arbitrarily high $E_{\rm cm}$ is therefore $L\approx L_{H}$ or
$\Omega_{H}L\approx E$ for either of the two particles. For
$E_{1}=E_{2}=E$, we denote
$L_{H1}=L_{H2}=L_{H}=\frac{2(n^2+r_H)}{a}$ and
Eq.~(\ref{eq:general_subextremal}) reduces to
\begin{equation}
\frac{E_{\rm cm}(r\to r_H)}{2\sqrt{m_1m_2}}=
\sqrt{\frac{(m_1-m_2)^2}{4m_1m_2}+1+\frac{(L_{1}-L_{2})^{2}L_{H}}
{4(L_{H}-L_{1})(L_{H}-L_{2})a}}. \label{eq:same_energy_subextremal}
\end{equation}
When NUT charge varnish ($n=0$), the above equation  reduces to the
result in Ref.~\cite{Grib_Pavlov20102052} for the kerr black hole.
As mention in Ref. \cite{WLGF2010}, to obtain an arbitrarily high CM
energy $E_{\rm cm}$, one of the colliding particles should have
critical angular momentum $L_H$. We assume that $L_1\to L_H$ and
obtain
\begin{equation}
\frac{E_{\rm cm}}{2\sqrt{m_1m_2}}
\simeq\sqrt{\frac{(L_{2}-L_{H})L_H}{4(L_1-L_{H})a}}.
\label{eq:upper_bound_marginally_bound}
\end{equation}
Now we will discuss how the rotating parameter $a$, and the NUT
charge $n$ affect the CM energy $E_{\rm cm}$. We denote the small
parameter $\xi=a_{\text{max}}-a$ with
$a_{\text{max}}=\sqrt{1+n^{2}}$. For fixed NUT charge $n$ and $\xi$,
the range $[L_{\text{min}},\;L_{\text{max}}]$ of angular momentum
for the particles to reach the horizon can be determined numerically
with the effective potential $V_{\rm{eff}}$ for the near-extremal KNT
spacetime.
\begin{table}[h]
\begin{center}
\caption{The CM energy per unit rest mass $\frac{E_{\rm{cm}}}{m_1}$
for the KTN spacetime with rotating parameter
$a=a_{\text{max}}-\xi$, $m_1=m_2=1$ and $L_{1}=L_{\text{max}}$,
$L_{2}=L_{\text{min}}$.}\label{TableCMenergy}
\begin{tabular}{c c c c c c }
  \hline
  \hline
   & $\xi$=0.1 & $\xi$=0.01 & $\xi$=0.001 & $\xi$=0.0001 \\
  \hline $n$=0\;\; &   6.901  & 12.5354 & 22.6352 & 40.4856 \\
 $n$=0.2  & 6.8814 & 12.4733 & 22.2955 & 39.6429 \\
 $n$=0.4  & 6.8248 & 12.1373 & 21.4312 & 38.2908 \\
 $n$=0.6  & 6.7369 & 11.7017 & 19.8102 & 36.3196 \\
 $n$=0.8  & 6.6546 & 11.3053 & 19.1093 & 33.1950 \\
  \hline
\end{tabular}
\end{center}
\end{table}
For a angular momentum $L\in[L_{\text{min}},\;L_{\text{max}}]$, we
can get a negative $V_{\rm{eff}}(l)$ for $r>r_{H}$. However, for
arbitrary charge $n$ and $\xi$, we find that, within a small range
near the horizon $r_{H}$, the effective potential
$V_{\rm{eff}}(L_{H})$ is always positive, which means the angular
momentum $L_{H}$ does not lie in the range
$[L_{\text{min}},\;L_{\text{max}}]$. So the CM energy $E_{\rm{cm}}$
in (\ref{eq:same_energy_subextremal}) is not divergent. Thus, the CM
energy is finite for arbitrary charge $n$ and rotating parameter
$a$. Considering that one of the colliding particles has the maximum
angular momentum $L_{\text{max}}$ and another one has the minimum
angular momentum $L_{\text{min}}$, we obtain the CM energy per unit
rest mass for different $n$ and $\xi$. The result is shown in Table
\ref{TableCMenergy}. From the table, we can see that, for the KTN
spacetime with rotating parameter $a$ less than $a_{\text{max}}$
there will be an upper bound for the CM energy. It is also suggested
that the CM energy grows very slowly as the maximally spinning case
($\xi\rightarrow 0$) is approached. For fixed parameter $\xi$, the
value of CM energy decreases with the increase of the charge $n$.
For the case $n=0$, it reduces to result for the Kerr black hole and
we recover the numerical result in Ref.
\cite{Jacobson_Sotiriou2010}.

\subsection{Near-horizon collision in the extremal KTN spacetime}
\label{subsec:extreme}

For the extremal KTN spacetime, the rotating parameter $a$ and NUT
charge $n$ satisfy the relation $n^2=a^2-1$, the numerator
$A(r)-B(r)$ and the denominator $C(r)$ both must have a second-order
zero at $r=r_{H}$. Using l'Hospital's rule twice, we obtain
\begin{equation}
\frac{E_{\rm cm}(r\to r_H)}{2\sqrt{m_1m_2}}
=\sqrt{\frac{(m_1-m_2)^2}{4m_1m_2}+\frac{A(r)''-B(r)''}{C(r)''}}\bigg|_{r=r_H},
\end{equation}
where the second-order derivatives are given by
\begin{eqnarray*}
A''(r)|_{r=r_H}&=&-2L_1L_2+2(1+3E_1E_2)n^2+2r_H(-4+5r_H+E_1E_2(2+5r_H))\\\nonumber
b''_{i}(r)|_{r=r_H}&=&-2L_i^2+2(3E_i^2-1)n^2+2r_H(4-5r_H+E_i^2(2+5r_H))\\\nonumber
C''(r)|_{r=r_H}&=&2(n^2+r_H(5r_H+4)) \\
B''(r)|_{r=r_H}&=&B\left[\frac{1}{2}\left(
\frac{b_{1}''}{b_{1}}+\frac{b_{2}''}{b_{2}}\right)
-\frac{1}{4}\left(\frac{b_{1}'}{b_{1}}-\frac{b_{2}'}{b_{2}}\right)^{2}
\right].
\end{eqnarray*}
After tedious calculation, we obtain the CM energy at the outer
horizon for the extremal KTN spacetime
\begin{equation}
\frac{E_{\rm cm}(r\to r_{H})}{2\sqrt{m_1m_2}}=
\sqrt{\frac{(m_1-m_2)^2}{4m_1m_2}+1+\frac{[(L_{H1}-L_{1})
-(L_{H2}-L_{2})]^{2}+\frac{(L_{H1}L_2-L_{H2}L_1)^2}{4a^2}
}{4(L_{H1}-l_{1})(L_{H2}-L_{2})}}. \label{eq:general_subextremalaaa}
\end{equation}
For the case $E_1=E_2=1$, Eq. (\ref{eq:general_subextremalaaa})
reduces to
\begin{equation}
\frac{E_{\rm cm}(r\to r_{H})}{2\sqrt{m_1m_2}}=
\sqrt{\frac{(m_1-m_2)^2}{4m_1m_2}+\frac{1}{2}\Big(\frac{L_1-\check{L}_{H}
}{L_2-\check{L}_H}+\frac{L_2-\check{L}_H}{L1-\check{L}_H}\Big)}.
\label{eq:general_subextremalddd}
\end{equation}
where the critical angular momentum $\check{L}_H=2a$. For the
special case $m_1=m_2$ and $n=0$, the above equation recovers the
result obtained by BSW \cite{BSW2009}. The expression
(\ref{eq:general_subextremalddd}) shows that the unlimited CM energy
can be approached if one of the colliding particles has critical
angular momentum $\check{L}_H$ which ensures the particle can reach
the outer horizon. Since the particles move along the stable
circular orbits in the equatorial plane, there must exist a
restriction for the angular momentum, which is also shown in Table
\ref{TableCMenergy11}.
\begin{table}[h]
\begin{center}
\caption{The ranged angular momentum $L$ for the extremal KTN spacetime
 with different rotating parameter $a$ and NUT charge $n$.
}\label{TableCMenergy11}
\begin{tabular}{c c c c c c }
  \hline
  \hline
      & $a$=1 & $a$=1.1 & $a$=1.2 & $a$=$\sqrt{2}$ & $a$=1.8 \\
   \hline $L_{\text{max}}$&2\;& 2.048  & 2.350 & 2.828 & 3.536 \\
 $L_{\text{min}}$&-4.828  & -5.027 & -5.222 & -5.633 & -6.353 \\
  \hline
\end{tabular}
\end{center}
\end{table}

With the help the effective potential
(\ref{eq:effective_potential}), we can determine the range of the
rotating parameter $a$ for the extremal KTN spacetime. The
effective potential for a particle with critical angular momentum
$\check{L}_H$ is
\begin{eqnarray}\label{rc}
 V_{\rm{eff}(L\to \check{L}_H}) =-\frac{(r-1)^2 \left(r+1-a^2\right)}
 {(r^2+a^2-1)^2}.
\end{eqnarray}
As expected, the effective potential $ V_{\rm{eff}(L\to
\check{L}_H)}$ approaches 0 at spatial infinity. Obviously, the
condition for the particle falling freely from rest at infinity to
the horizon can be expressed as
\begin{eqnarray}
  V_{\rm{eff}(L\to \check{L}_H)} \leq 0 \quad \mbox{for any}\quad  r\geq
  1.
\end{eqnarray}
Combing with the condition $1+n^2\geq a^2$, we can solve Eq.
(\ref{rc}) and obtain the range for the parameters $a$ and $n$
\begin{eqnarray}
 1\leq a\leq \sqrt{2}, ~~~ |n|\leq1,
\end{eqnarray}
which means that for the KTN extremal spacetime with $a \in
[1,\;\sqrt{2}]$ and the NUT charge $|n|\leq1$, the particle with
critical angular momentum $\check{L}_H$ can reach the horizon. Thus, for the fixed $a\in[1,\;\sqrt{2}]$, one can find
that the CM energy will be unlimited if $L_1=\check{L}_{H}$ and
$L_{2}$ is in a proper range. In Fig.1, we plot the effective
potential $V_{\rm{eff}(L\to \check{L}_H)}$ and CM energy
$E_{\rm{cm}}$ of collision for the different values of $a$ and $n$.
From Fig. 1 (a), one can find that for $1\leq a\leq \sqrt{2}$ the
effective potential $V_{\rm{eff}(L\to \check{L}_H)}$ is negative
when $r\geq r_H=1$ so that the particle can reach the horizon.
However,  the effective potential $V_{\rm{eff}(L\to \check{L}_H)}$
for $a=1.8$ is positive near the horizon, which implies that the
particle can not reach the horizon in this case. From Fig.1 (b), we
also find that for the case $1\leq a\leq\sqrt{2}$ the CM energy at
the horizon is be unlimited, which can be explained by a fact that
the effective potential $V_{\rm{eff}(L\to \check{L}_H)}$ is negative
for the two colliding particles with angular momenta
$L_{1}=\check{L}_H$ and $L_{2}=-2$. For the case $a=1.8$, it is
obvious that the CM energy is limited and the particle cannot reach
the horizon. Moreover, we also find that with the increase of the
NUT charge $n$ the CM energy increases, but the rate of increase of
the CM energy decreases in the KTN extremal spacetime.

\begin{figure}[ht]
\begin{center}\label{a1111}
\includegraphics[width=8cm]{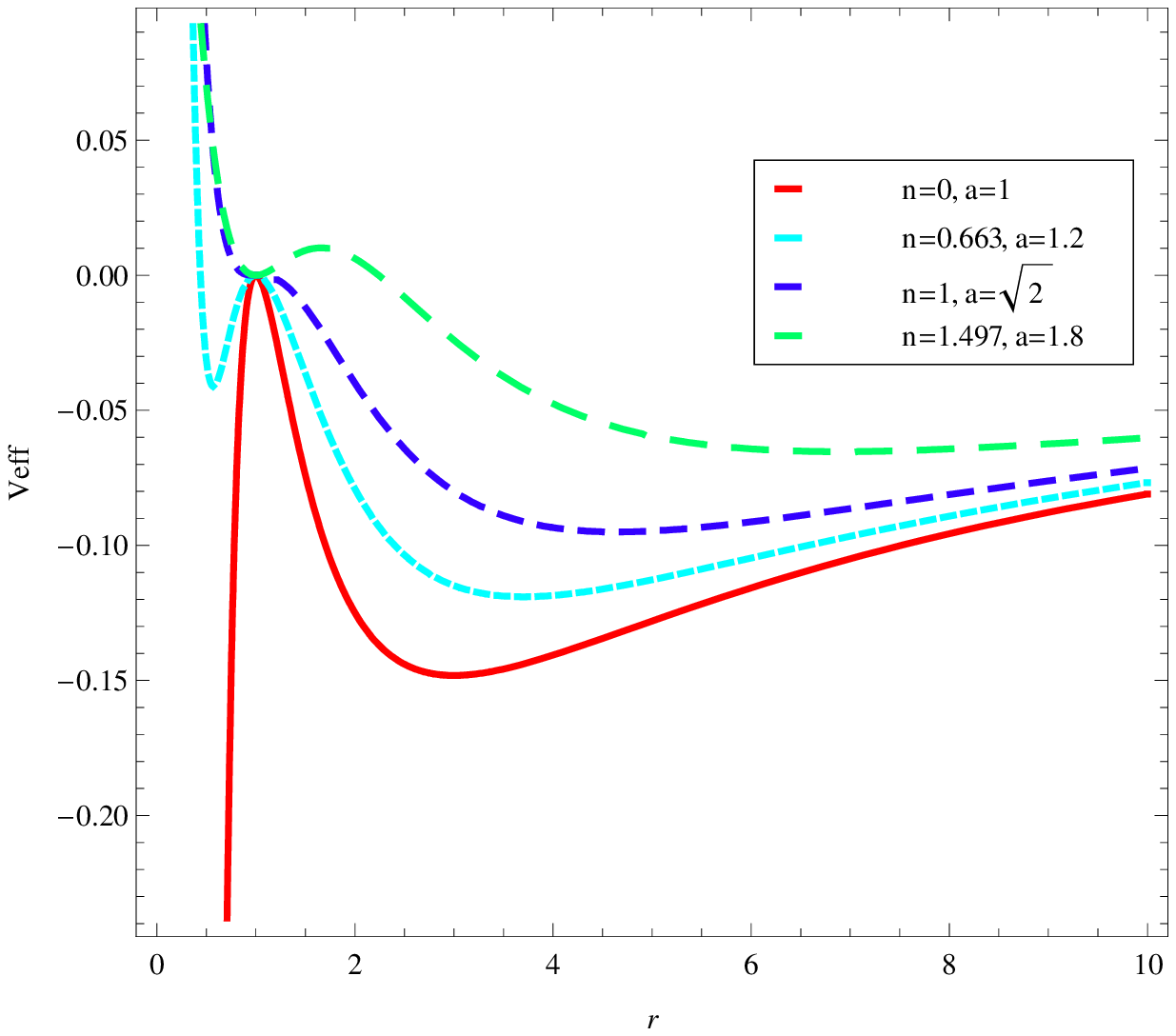}
\includegraphics[width=8cm]{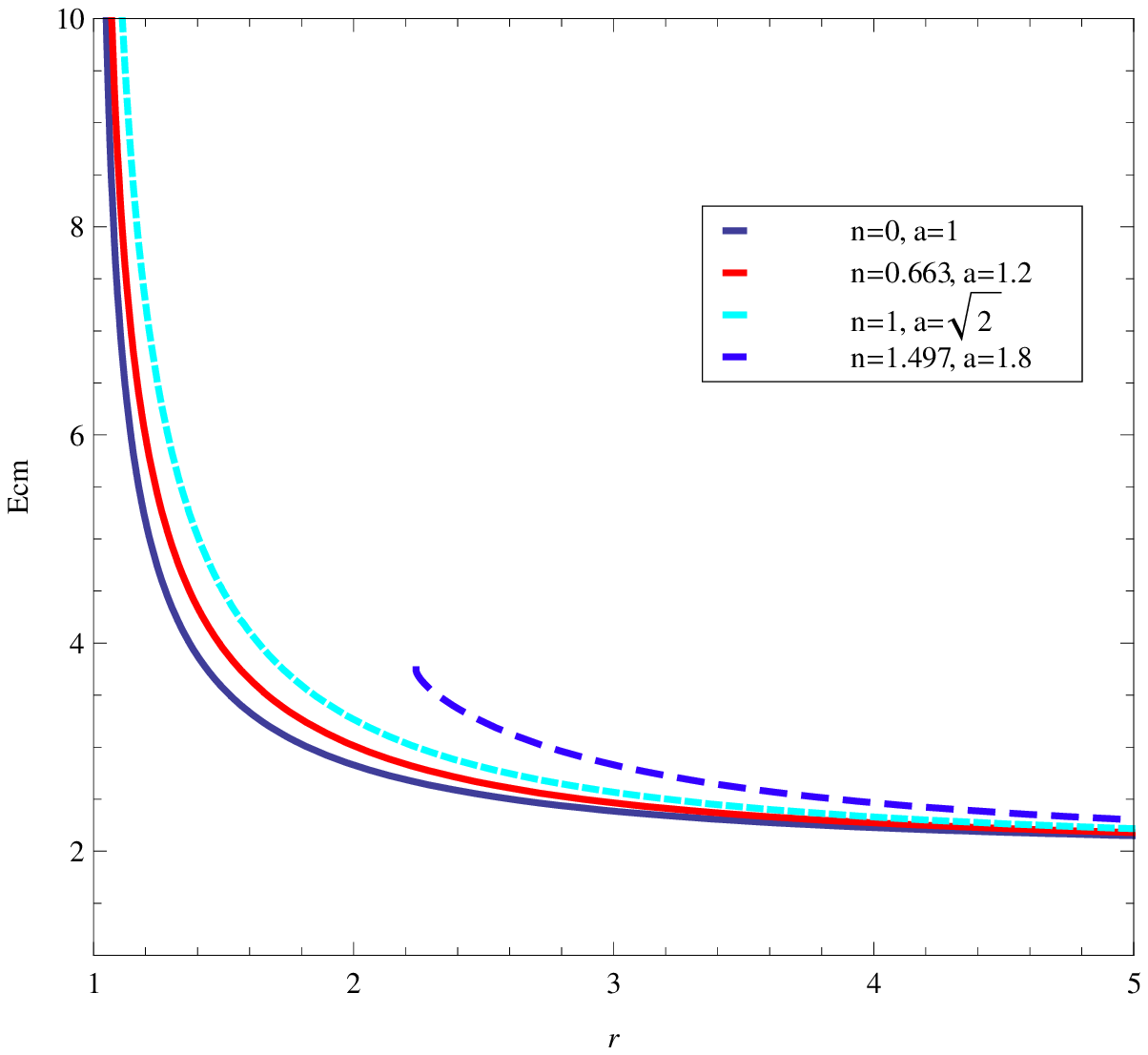}
\caption{(Color online) For an extremal  KTN spacetime (a) the
variation of the effective potential $V_{\rm{eff}(L\to
\check{L}_H)}$ with radius vs $a$ with angular momentum
$L=\check{L}_H=2a$. (b) the variation of the CM energy $E_{\rm{cm}}$
with radius vs $a$ with $m_1=m_2=1$ and $L_{1}=\check{L}_H$,
$l_2=-2$.}
\end{center}
\label{fig1}
\end{figure}

Now we would like to estimate the maximal value of the CM energy of
the collision particles in the background of the  extremal KTN spacetime.
Here, we consider the case the angular momentum of one of the
particles deviates little from the critical angular momentum
$\check{L}_H$. For simplicity, we choose $L_{1}=\check{L}_H-\delta
L$ and $L_{2}=0$. From Eq. (\ref{eq:general_subextremalddd}), we
obtain that the maximal CM energy for the collision particles can be
approximated as
\begin{eqnarray}
 \frac{E_{\rm cm}^{max}}{\sqrt{m_1m_2}}\approx \sqrt{2a}
 \delta L^{-1/2}+O(\delta L^{1/2})=\sqrt{2\sqrt{1+n^2}}
 \delta L^{-1/2}+O(\delta L^{1/2}) ,
\end{eqnarray}
in the extremal KTN spacetime. Clearly, the maximal CM energy
increases  with the increase of NUT charge. Now, we also estimate
the maximal value of the CM energy of particles in the case of the
near-extremal case. Here  we denote a small deviation
 $\xi=a_{\text{max}}-a \ll 1$, and suppose that
$L_{1}=\check{L}_H$ and $L_{2}=0$. Then, it is easy to obtain from
Eq. (\ref{eq:same_energy_subextremal}) that the maximal CM energy of
the collision particles can be estimated by
\begin{eqnarray}
 \frac{E_{\rm cm}^{max}}{\sqrt{m_1m_2}}\approx 2\sqrt[4]
 {\frac{a^3}{8}}\delta \xi^{-1/4}+O(\delta \xi^{1/4}),
\end{eqnarray}
in the near-extremal KTN spacetime. This means that the
maximal CM energy of the collision particles increases
proportionally to $a^{3/4}$. We assume that  the rest masses of the
colliding particles $m_1$, $m_2$ are of about 1 GeV, just like the
mass of a neutron. In order to obtain the Planck-scale energy
$E_{\rm cm}\sim 10^{19}$ GeV, we need $\delta \xi \sim 10^{-76}$,
which is similar to that in the Kerr-Newman black hole
\cite{WLGF2010}. This implies that it is very hard for a
near-extremal case to be a particle accelerator of
Planck-scale energy.

\section{summary}

In this paper, we studied the collision of two particles  with the
different rest masses moving in the equatorial plane of the KTN spacetime
and get the center-of-mass (CM) energy for the particles. Our
result shows that the CM energy depends not only on the rotation
parameter of the KTN spacetime, $a$, but also on the NUT charge of the
KTN spacetime, $n$. For the extremal KTN spacetime, the presence of
the NUT charge modified the restrict conditions for the spin $a$
when arbitrarily high CM energy appears, i.e., $1\leq
a\leq\sqrt{2}$, which is a significant difference from the Kerr
\cite{BSW2009} and Kerr-Newman \cite{WLGF2010} black holes. For the
near-extremal case, we also found  that the CM energy $E_{\rm
cm}$ decreases with the increase of the NUT charge $n$ when one
particle has the maximum angular momentum $L_{\text{max}}$ and the
other has the minimum angular momentum $L_{\text{min}}$. We also
estimated the maximal value of the CM energy of particles  for both
the extremal and the near-extremal KTN spacetime when one particle has
the critical angular momentum and the other has zero angular
momentum and discussed the change of the maximal CM energy with the
parameters $a$ and $n$.

\begin{acknowledgments}

This work was supported by the National Natural Science Foundation
of China under Grant No 10875040, 10875041;  the key project of the
National Natural Science Foundation of China under Grant No
10935013;  the National Basic Research of China under Grant No.
2010CB833004,   PCSIRT under Grant No.  IRT0964,
and the Construct Program  of the National Key Discipline.

\end{acknowledgments}

\end{document}